\documentclass[journal]{IEEEtran}
\ifCLASSINFOpdf
\usepackage[pdftex]{graphicx}
\else
\fi
\usepackage{amssymb}
\usepackage{bm}
\usepackage{color}
\usepackage{textcomp}

\usepackage{ulem}

\usepackage{graphicx}
\usepackage{subfigure}
\usepackage{dsfont}
\usepackage{mathtools}
\usepackage{algpseudocode}
\usepackage{algorithmicx,algorithm}

\usepackage[caption=false,font=footnotesize]{subfig}
\begin{document}
%
\title{Auxiliary Diagnosing Coronary Stenosis Using Machine Learning}
%
%
%

\author{Weijun ZHU (1)(2), Fengyuan LU (1), Xiaoyu YANG *(1) and En LI (1)
	
	(1) the Second Affiliated Hospital, Zhengzhou University, Zhengzhou, China.
	
	(2) School of Information Engineering, Zhengzhou University, Zhengzhou, China.

    * Corresponding author: Xiaoyu YANG}
\maketitle


\begin{abstract}
How to accurately classify and diagnose whether an individual has Coronary Stenosis (CS) without invasive physical examination? This problem has not been solved satisfactorily. To this end, the four machine learning (ML) algorithms, i.e., Boosted Tree (BT), Decision Tree (DT), Logistic Regression (LR) and Random Forest (RF) are employed in this paper. First, eleven features including basic information of an individual, symptoms and results of routine physical examination are selected, as well as one label is specified, indicating whether an individual suffers from different severity of coronary artery stenosis or not. On the basis of it, a sample set is constructed. Second, each of these four ML algorithms learns from the sample set to obtain the corresponding optimal classified results, respectively. The experimental results show that: RF performs better than other three algorithms, and the former algorithm classifies whether an individual has CS with an accuracy of 95.7\% (=90/94).
\end{abstract}


%
\IEEEpeerreviewmaketitle

\section{Introduction}\label{1}
%
%
%
%
\IEEEPARstart{T}{he} World Health Organization believes that cardiovascular diseases (CVDs) have become the number one killer threatening human life [1]. In 2016, 17.9 million people died of such diseases, accounting for 31\% of the total global deaths in that year [1]. In fact, CVDs includes a number of different specific diseases, and Coronary Artery Disease (CAD) is one of the most fatal CVDs [2].

Coronary stenosis forms the essence of CAD. Unfortunately, the incidence rate of CAD/CS is increasing at present. Its incidence is related to many risk factors. The typical symptoms of patients are compression pain in anterior chest area accompanied by tightness and dyspnea. And the symptoms can aggravate after activity and reduce after rest. However, the difference of symptoms is very large in clinical practice. Some patients feel discomfort and pain with left shoulder as the first symptom. Others feel epigastric discomfort, nausea, neck tightening, or acupuncture pain in chest, lasting only a few seconds. Someone only feels toothache, foot pain and so on. It is obviously that the symptoms are changeable. The diagnosis of coronary heart disease is a relatively complex process, which requires professional doctors to analyze the symptoms, risk factors, ECG, Holter, cardiac ultrasound, myocardial enzymology, coronary CTA, coronary angiography and so on.

Coronary angiography has long been recognized as the gold standard for the diagnosis. However, it is difficult for doctors in grassroots hospitals and community hospitals to conduct coronary angiography. In addition, inserting a catheter into an artery is, after all, an invasive physical examination, not suitable for large-scale screening in the general population. Furthermore, coronary angiography may lead to some complicating diseases [3], even death in rare cases [4][5].

Thus, we are looking for a way to assist in the CS diagnosis in safety and screen of coronary heart disease, as well as control the false negative rate to a certain range, considering that a false negative diagnosis is more dangerous than a false positive one. Can an AI algorithm be used to do something for this? We need an AI-based method. With this method, we input the data including the basic information of an individual, routine physical examination and symptoms, and the AI-based method can intelligently and automatically analyze these data, so that the relationship between these raw data and the diagnosis conclusion made by coronary angiography are obtained. In this way, the result of coronary angiography can be predicted, indicating whether an individual has coronary stenosis. To this end, the four machine learning algorithms are employed respectively in this study. Thus, a powerful ML-based approach for auxiliary diagnosing coronary stenosis is proposed. This is the contribution of this study.

\section{Objective}\label{2}

We will explore the ability and efficiency of each of the several machine learning algorithms in terms of the intelligent qualitative diagnosis for coronary artery stenosis, according to an individual personal basic information, routine physical examination and symptoms. 

Given a sample data set consisting of individuals’ data, what is the highest accuracy we can get, by adjusting values of hyper-meters (also including the division of training set and testing set) on this data set? And which ML algorithm can get this best accuracy? We will try to answer these questions, with the following method and experiments.

\section{Method}\label{2}

A new method is designed, as shown in Fig.1. In this study, the following four ML algorithms have been already employed, respectively: BT [6], DT [7], LR [8] and RF [9]. In a word, some data will be inputted to generate a ML model, which can be used to predict the results of coronary angiography (classify the level of coronary stenosis) for other data. Section IV will explain the meaning of the data.

\section{The Raw Data and the Experimental Platform}\label{2}

All the raw data used in this study are originated from the actual clinical data. One thousand records (samples) are selected randomly. Training and classification will be performed on this sample set which has one thousand samples (the raw dataset A), where: the value of label of 623 records is 1, which means ”severe coronary stenosis”; the value of label of 125 records is 2, which means ”coronary artery is moderate or mild stenosis”; the value of label of 252 records is 3, which means ”coronary artery is normal”. In other words, the raw data needed processed by the machine learning algorithms have a label of three classification. 

In this study, the factors influencing the diagnosis are selected as follows: gender, age, fasting plasma glucose (FPG), LDL, history of hypertension (how many years), history of diabetes (how many years), smoking history (how many years), sweating at the onset of the disease, ECG (ST segment is elevated, ischemic change occurs, or the above situation does not occur), whether a cardiac color Doppler ultrasound indicates plaques occur in the neck vessels or not, whether a cardiac color Doppler ultrasound indicates some abnormal wall motions or not (UCG). In other words, the machine learning algorithms need to process some raw data with the above eleven features. 

Fig.2 illustrates the relationship between each of the eleven features and the label, as well as the numbers of samples in the raw dataset. In every subfigure, X-axis means value of the corresponding one feature, and Y-axis means value of the label, as well as the color of a spot indicates the number of records when a value of corresponding feature and a value of label are given. The darker the color, the more records exist. Obviously, the raw data is balanced. It should be noted that, the values of the first four features are renormalized in these subfigures since such a renormalization is needed before ML training is performed, and Table I provides the relationship between them. 

The used experimental platform is as follows. 

(1) CPU: Intel i9-10900 @ 2.80GHz

(2) Memory: 32GB

(3) GPU: none

(4) Python, Anaconda [10] and Turicreate [11] are used to implement the machine learning algorithms, and conduct our experiments.

\section{Results}\label{2}
\subsection{Ability of Classification about an Individuals CS level (Three Classification)}\label{2.2}

The corresponding optimal classified results are obtained by adjusting the value of the hyper-parameters, using the four machine learning algorithms, respectively, as shown in Table II. This table also depicts the corresponding values of hyper-parameters which make the optimal classified results appear. Obviously, RF has the higher accuracy, reaching 92.6\%. Thus, RF is the preferred algorithm. In addition, Fig.3 shows the obtained confusion matrix.

\subsection{Ability of Classification about whether an Individual has CS or not (Binary Classification 1)}\label{2.2}

According to the confusion matrix shown in Fig.3, one can compute some values of metrics in terms of binary classification 1. RF shows an excellent performance. And its classified accuracy reaches (75+12+3)/(75+12+2+3+2)=90/94=95.7\%.

False negative rate (FNR) is the rate of missed diagnosis, and false positive rate (FPR) is the rate of misdiagnosis rate. Especially for missed diagnosis, it is very harmful to patients. FNR is controlled within 2/(2+75+3)=2/80=2.5\% in the case of binary classification 1 with RF, prompting that RF is a relatively suitable algorithm for our mission again. Table III depicts more values of some popular metrics in this situation.

\subsection{Ability of Classification about whether an Individual has a severe CS or not (Binary Classification 2)}\label{2.2}

According to the confusion matrix shown in Fig.3, one can compute some values of metrics in terms of binary classification 2. This time, RF shows an excellent performance again. Its classified accuracy reaches 87/94=92.6\%. And FNR is 2/(2+75)=2/77=2.6\% in the case of binary classification 2 with RF, prompting that RF is a suitable algorithm for our mission again. Table III depicts more values of some popular metrics in this situation.

\subsection{Efficiency and Summary}\label{2.2}

In our experiment, each of BT, DT, LR and RF runs quickly. The average running time of each algorithm for classifying one sample (one individual) is not more than 0.00008 seconds, as depicted in Table II. Furthermore, no significant difference in terms of the running speed occurs, among these four machine learning algorithms.

In summary, RF is more suitable than the other three ML algorithms, in terms of predicting the results of coronary angiography or classifying the level of coronary artery stenosis for an individual. 

\section{The related works}\label{2}
Some approaches directly use machine learning to diagnose the different levels of CS [12][13][14]. Table IV summarizes the advantage of the newly proposed method.

\section{Conclusions}\label{2}

In this study, the machine learning technique is employed to auxiliary diagnose coronary artery stenosis or predict the result of coronary angiography for individuals. To the best of our knowledge, this is the first AI-based approach which can directly decide whether or not an individual is suffering from the severe coronary stenosis (indicating a stent is needed).

Considering coronary angiography puts forward a high requirement for doctors and hospitals, the new RF-based method which is a noninvasive and easy-to-use solution, can be expected to be more suitable for large scale screening of CS in grass-roots hospitals. Considering that this disease is very fatal and dangerous, and the number of patients and the number of potential risk groups are very large, the comparative advantages and potential clinical prospects of the new method are obvious.



\begin{table*}[!t]
\newcommand{\tabincell}[2]{\begin{tabular}{@{}#1@{}}#2\end{tabular}}
\renewcommand{\arraystretch}{1.3}
\caption{The mapping relation between values of X-axis in the first four subfigures of Fig.2 and the true value of the first four features in Fig.1}
\label{table_1}
\centering
\begin{tabular}{cccc}
\hline
The subfigure & The feature & The value of feature & The value in subfigure\\
\hline
1 & gender & female & 0\\
\hline
1 & gender & male & 1\\
\hline
2 & age & 0-4 & 1\\
\hline
2 & age & 5-9 & 2\\
\hline
2 & age & 10-14 & 3\\
\hline
2 & age & 15-19 & 4\\
\hline
2 & age & 20-24 & 5\\
\hline
2 & age & 25-29 & 6\\
\hline
2 & age & 30-34 & 7\\
\hline
2 & age & 35-39 & 8\\
\hline
2 & age & 40-44 & 9\\
\hline
2 & age & 45-49 & 10\\
\hline
2 & age & 50-54 & 11\\
\hline
2 & age & 55-59 & 12\\
\hline
2 & age & 60-64 & 13\\
\hline
2 & age & 65-69 & 14\\
\hline
2 & age & 70-74 & 15\\
\hline
2 & age & 75-79 & 16\\
\hline
2 & age & 80-84 & 17\\
\hline
2 & age & 85-89 & 18\\
\hline
2 & age & >=90 & 19\\
\hline
3 & FPG & 0-1.9 & 0\\
\hline
3 & FPG & 1.9-2.9 & 1\\
\hline
3 & FPG & 2.9-3.9 & 2\\
\hline
3 & FPG & 3.9-5.6 & 3\\
\hline
3 & FPG & 5.6-6.1 & 4\\
\hline
3 & FPG & 6.1-6.7 & 5\\
\hline
3 & FPG & 6.7-7.7 & 6\\
\hline
3 & FPG & 7.7-8.7 & 7\\
\hline
3 & FPG & 8.7-9.7 & 8\\
\hline
3 & FPG & 9.7-10.7 & 9\\
\hline
3 & FPG & 10.7-11.1 & 10\\
\hline
3 & FPG & >11.1 & 11\\
\hline
4 & LDL & 0-1.07 & 0\\
\hline
4 & LDL & 1.07-2.07 & 1\\
\hline
4 & LDL & 2.07 ~ 3.37 & 2\\
\hline
4 & LDL & 3.37-4.12 & 3\\
\hline
4 & LDL & 4.12-4.89 & 4\\
\hline
4 & LDL & 4.89-6.89 & 5\\
\hline
4 & LDL & 6.89-8.89 & 6\\
\hline
4 & LDL & >8.89 & 7\\
\hline
\end{tabular}
\end{table*}

\begin{table*}[!t]
	\newcommand{\tabincell}[2]{\begin{tabular}{@{}#1@{}}#2\end{tabular}}
	\renewcommand{\arraystretch}{1.3}
	\caption{Optimal results and the corresponding values of hyper-parameters, when the four ML algorithms are employed on dataset A, respectively}
	\label{table_2}
	\centering
	\begin{tabular}{ccccc}
		\hline
		------ & RF & BT & DT & LR\\
		\hline
		optimal accuracy & 0.926 & 0.872 & 0.894 & 0.872\\
		\hline
		AUC & 0.803 & 0.748 & 0.755 & 0.801\\
		\hline
		average classified time for one sample (second) & 0.000078 & 0.000074 & 0.000076 & 0.000068\\
		\hline
		fraction in data.random\_split & 0.89 & 0.89 & 0.89 & 0.89\\
		\hline
		seed in data.random\_split & 2129 & 2129 & 2129 & 2129\\
		\hline
		seed in classifier & 15073 & 292 & 603 & 12\\
		\hline
		random\_seed in classifier & 2491 & --- & --- & ---\\
		\hline
	\end{tabular}
\end{table*}

\begin{table*}[!t]
	\newcommand{\tabincell}[2]{\begin{tabular}{@{}#1@{}}#2\end{tabular}}
	\renewcommand{\arraystretch}{1.3}
	\caption{some values of metrics in terms of binary classification}
	
	\label{table_3}
	\centering
	\begin{tabular}{cccccc}
		\hline
		------ & TPR & FPR & TNR & FNR & accuracy\\
		\hline
		binary classification 1 (classifying CS) & 0.975 & 0.143 & 0.857 & 0.025 & 0.957\\
		\hline
		binary classification 2 (classifying severe CS) & 0.974 & 0.294 & 0.706 & 0.026 & 0.926\\
		\hline
	\end{tabular}
\end{table*}

\begin{figure*}
	\centering
	\scalebox{1}{\includegraphics[width=0.5\textwidth]{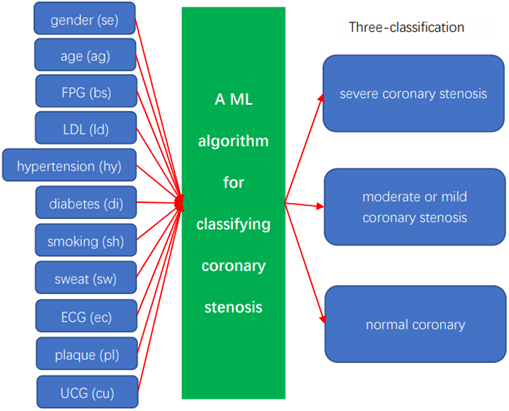}}%
	\caption{{\normalsize A novel method for auxiliary diagnosing coronary stenosis using machine learning}}
	\label{fig_1}
\end{figure*} 

\section*{Acknowledgment}

This work has been supported by the National Natural Science Foundation of China under Grant U1204608.

\ifCLASSOPTIONcaptionsoff
  \newpage
\fi



%
\section*{References}

[1] "Cardiovascular diseases, https://www.who.int/newsroom/fact-sheets/detail/cardiovascular-diseases-(cvds), World Health Organization, 2017."

[2] "Types of cardiovascular disease, https://www.who.int/cardiovascular diseases/en/cvd atlas 01 types.pdf?ua=1, World Health Organization, 2017."

[3] "Coronary angiogram - Mayo Clinic, https://www.mayoclinic.org/tests-procedures/coronary-angiogram/about/pac-20384904, 2021"

[4] "Kennedy JW, Baxley WA, Bunnel IL, Gensini GG, Messer JV, Mudd JG, Noto TJ, Paulin S, Pichard AD, Sheldon WC, Cohen M. Mortality related to cardiac catheterization and angiography. Cathet Cardiovasc Diagn. 1982;8(4):323-40. doi: 10.1002/ccd.1810080402. PMID: 7127459."

[5] "Ercan S, Kaplan M, Aykent K, et al, Sudden death after normal coronary angiography and possible causes, Case Reports, 2013 : bcr2013008753, doi: 10.1136/bcr-2013-008753."

[6] "Boosted Trees Classifier, https://apple.github.io/turicreate/docs/userguide/supervised-learning/boosted\_trees\_classifier.html"

[7] "Decision Tree Classifier, https://apple.github.io/turicreate/docs/userguide/supervised-learning/decision\_tree\_classifier.html"

[8] "Logistic Regression, https://apple.github.io/turicreate/docs/userguide/supervised-learning/logistic-regression.html"

[9] "Random Forest Classifier, https://apple.github.io/turicreate/docs/userguide/supervised-learning/random\_forest\_classifier.html"

[10] "Anaconda | The World's Most Popular Data Science Platform, https://www.anaconda.com/"

[11] "Overview · GitBook, https://apple.github.io/turicreate/docs/userguide/"

[12] "Zhang H, Wang X, Liu, C, et al, A METHOD for DETECTING CORONARY ARTERY STENOSIS BASED on ECG SIGNALS, Journal of Mechanics in Medicine and Biology, v 21, n 1, February 2021. "

[13] "Cios K, Goodenday L, Merhi M, Neural networks in detection of coronary artery disease, Computers in Cardiology,  p 33-37, Sep 1989."

[14] "Alizadehsani R, Roshanzamir M, Abdar M, Model uncertainty quantification for diagnosis of each main coronary artery stenosis, Soft Computing, v 24, n 13,  p 10149-10160, July 1, 2020."

\begin{figure*}
	\centering
	\subfigure[gender (se)]{
		\includegraphics[width=0.3\textwidth]{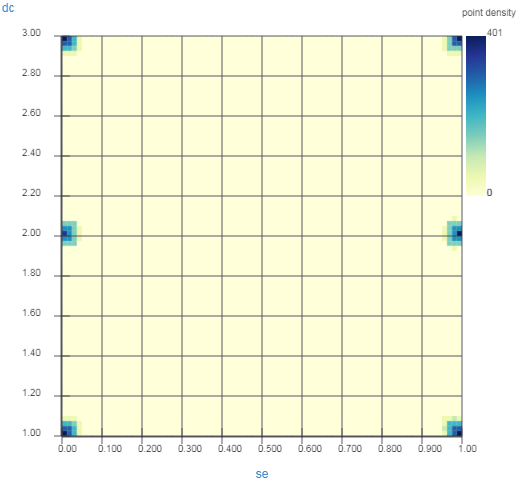}
	}
	\quad
	\subfigure[age (ag)]{
		\includegraphics[width=0.3\textwidth]{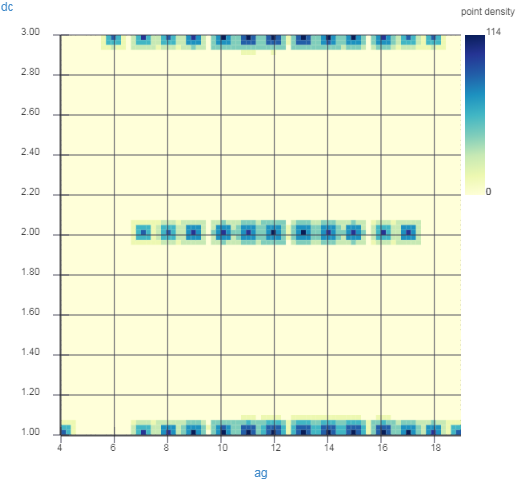}
	}
	\quad
	\subfigure[FPG (bs)]{
		\includegraphics[width=0.3\textwidth]{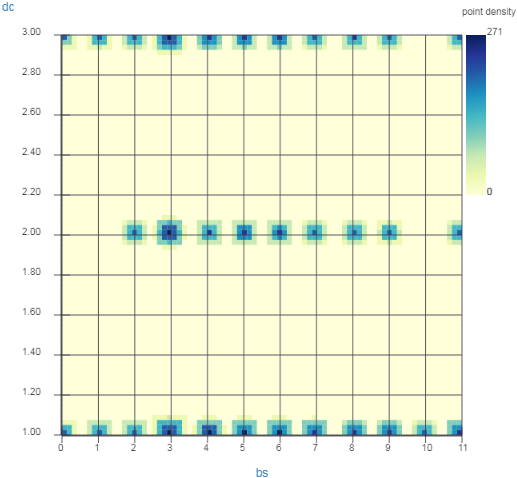}
	}
	\quad
	\subfigure[LDL (ld)]{
		\includegraphics[width=0.3\textwidth]{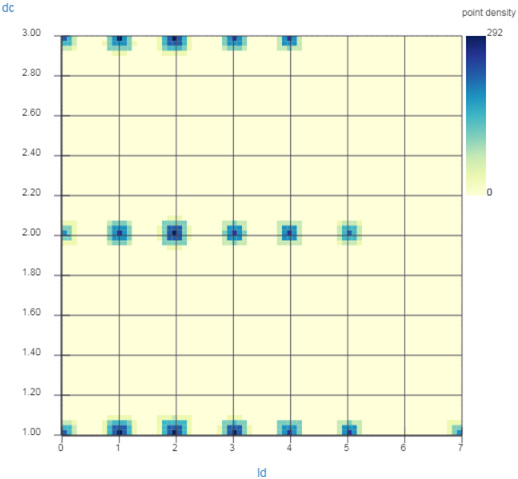}
	}
	\quad
	\subfigure[hypertension (hy)]{
		\includegraphics[width=0.3\textwidth]{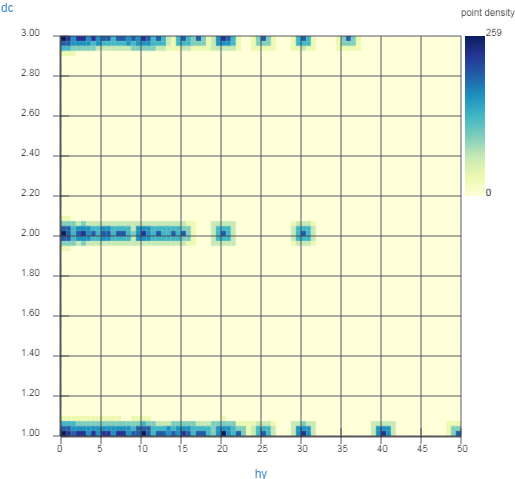}
	}
	\quad
	\subfigure[diabetes (di)]{
		\includegraphics[width=0.3\textwidth]{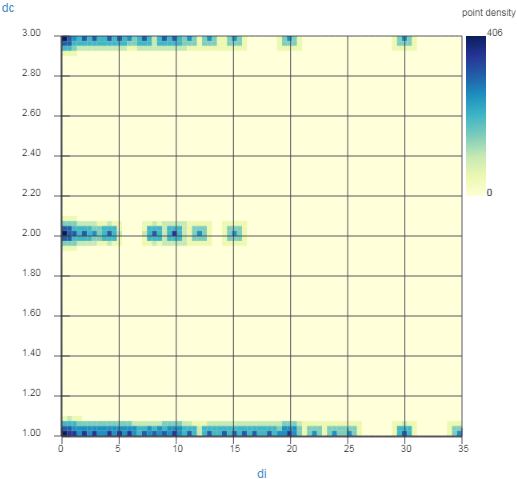}
	}
    \quad
	\subfigure[smoking (sh)]{
		\includegraphics[width=0.3\textwidth]{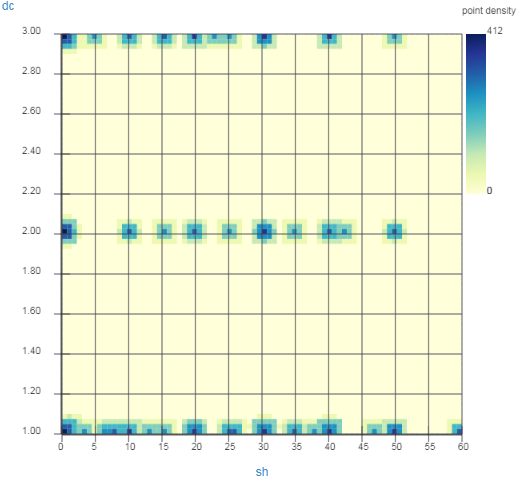}
	}
	\quad
	\subfigure[sweat (sw)]{
		\includegraphics[width=0.3\textwidth]{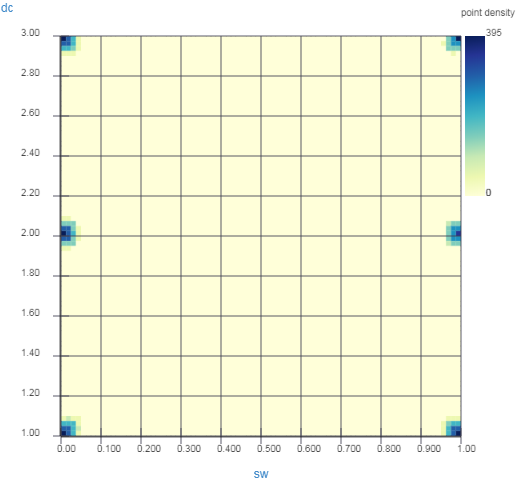}
	}
	\quad
	\subfigure[ECG (ec)]{
		\includegraphics[width=0.3\textwidth]{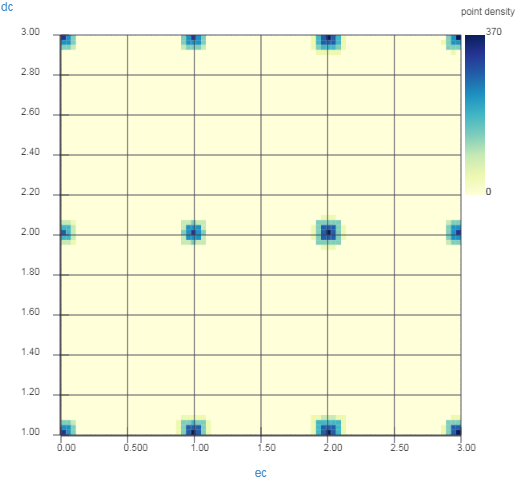}
	}
	\quad
	\subfigure[plaque (pl)]{
		\includegraphics[width=0.3\textwidth]{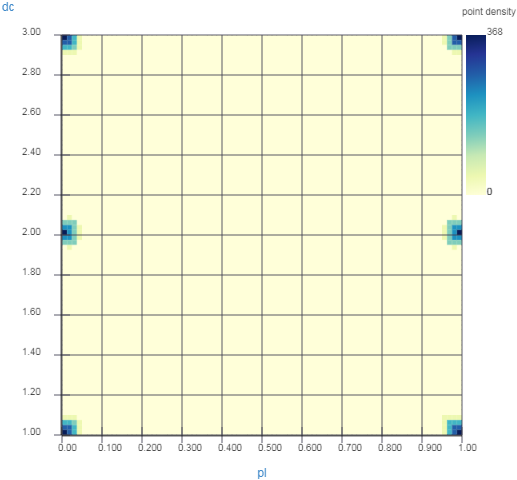}
	}
	\quad
	\subfigure[UCG (cu)]{
		\includegraphics[width=0.3\textwidth]{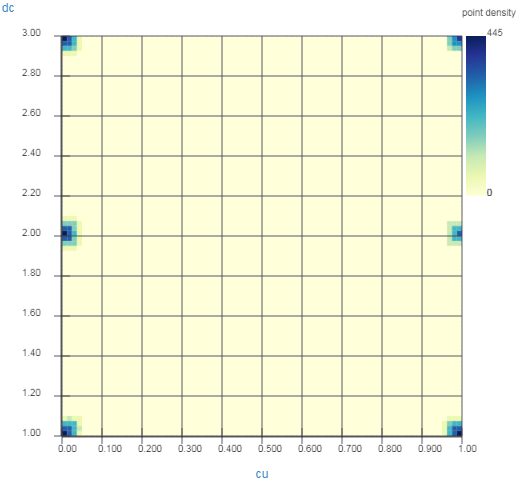}
	}
	\caption{{\normalsize each pair of 11 features \& 1 label, and its numbers of samples in raw data}}
\end{figure*}

\begin{figure*}
	\centering
	\scalebox{1}{\includegraphics[width=0.6\textwidth]{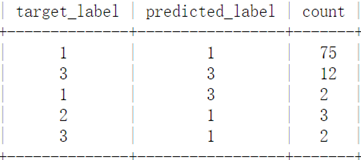}}%
	\caption{{\normalsize the obtained confusion matrix on A when RF is employed}}
	\label{fig_3}
\end{figure*} 

\begin{table*}[!t]
	\newcommand{\tabincell}[2]{\begin{tabular}{@{}#1@{}}#2\end{tabular}}
	\renewcommand{\arraystretch}{1.3}
	\caption{a comparison between the new method and the existing ones for CS diagnosis}
	
	\label{table_4}
	\centering
	\begin{tabular}{ccccc}
		\hline
		The methods & the one in [12] & the one in [13] & the one in [14] & The new one\\
		\hline
		number of samples & 200 & 65 & 300 & 1000\\
		\hline
		accuracy & 98.5\% (200 samples) & 66.6\% (65 samples) & 86.4\% (300 samples) & 100\% (whether 200 samples or 300 samples)\\
		\hline
		Can it classify the three different level of CS? & yes & no & no & yes\\
		\hline
	\end{tabular}
\end{table*}

%








\end{document}